\documentclass[12pt]{article}
\begin{document}
\begin{center}
\large\bf{Using the Renormalization Group}
\end{center}
\vspace{2cm}
\begin{center}
D.G.C. McKeon\\
Department of Applied Mathematics\\
University of Western Ontario\\
London  N6A 5B7  Canada\\
Tel: 519-661-2111, ext. 88789 \\
Fax: 519-661-3523\\
Email: dgmckeo2@uwo.ca
\end{center}

In computing quantum effects, it is necessary to perform a sum over all intermediate states consistent with prescribed initial and final states. Divergences arising in the course of evaluating this sum forces one to ``renormalize'' parameters characterizing the system. An ambiguity inherent in this rescaling is parameterized by a dimensionful parameter $\mu^2$ which serves to set a scale for the process. Requiring that the explicit and implicit dependence of a physical quantity on $\mu^2$ conspire to cancel leads to the so-called ``renormalization group'' equation [1-10]. It has proved possible to extract a lot of useful information from this equation; we will enumerate a number of these in this report\vspace{.3cm}.

\noindent
{\bf\large(I)}  The first instance we will analyze to show the utility of this approach is the relationship between the bare and renormalized parameters in the context of dimensional regularization and mass independent renormalization [4,11-13].

In this approach, bare quantities appearing in the $n = 4 - 2\epsilon$ dimensional Lagrangian are expanded in terms of poles at $\epsilon = 0$. In particular, the bare coupling $g_B$ and the renormalized coupling $g$ are related by 
$$g_B =\mu^\epsilon \sum_{\nu = 0}^\infty \frac{a_\nu (g)}{\epsilon^\nu}\eqno(1)$$
where $\mu$ is a renormalization induced scalar parameter. In minimal subtraction,
$$a_0 (g) = g.\eqno(2)$$
Explicit calculation leads to
$$a_\nu (g) = \sum_{k = \nu}^\infty a_{k,\nu}g^{2k+1}\eqno(3)$$
with 
$$a_{0,0} = 1.\eqno(4)$$
The sum in eq. (1) can now be reorganizaed to result in
$$g_B = \mu^\epsilon \sum_{k=0}^\infty g^{2k+1} S_k \left(\frac{g^2}{\epsilon}\right)\eqno(5)$$
where 
$$S_k (\xi) = \sum_{\ell = 0}^\infty a_{k + \ell , \ell}\;\xi^\ell \eqno(6)$$
and
$$S_k (0) = a_{k,0}.\eqno(7)$$
Since $g_B$ is independent of $\mu$,
$$\mu \frac{dg_B}{d\mu} = \left(\mu \frac{\partial}{\partial\mu} + \mu \frac{dg}{d\mu} \frac{\partial}{\partial g}\right) g_B = 0.
\eqno(8)$$
Equation (8) and (1) are consistent at order $\epsilon^0$ provided
$$ a_1 - \frac{a_0}{a_0^\prime} a_1^\prime + \beta a_0^\prime = 0 \eqno(9)$$
where
$$\mu \frac{dg}{d\mu} = \beta (g) - \epsilon \frac{a_0}{a_0^\prime} , \eqno(10)$$
so that
$$\beta (g) = \frac{a_0^2}{a_0^{\prime\,2}} \frac{d}{dg}\left(\frac{a_1}{a_0}\right).\eqno(11)$$
Terms of order $\epsilon^{-k}$ result in the equation
$$- \frac{a_0^2}{a_0^\prime} \frac{d}{dg}\left(\frac{a_k}{a_0}\right) + \beta a^\prime_{k-1} = 0\eqno(12)$$
so that $a_k$ is determined by $a_{k-1}$ and $\beta(k \geq 2)$, as well as $a_0(g)$. We also find that together eqs. (1) and (6) result in
$$\sum_{n=0}^\infty g^{2n} \left[ \frac{g^3}{\xi} S_n (\xi) + \left( - \frac{a_0}{a_0^\prime} \frac{g^2}{\xi} + \beta\right)\left((2n+1) S_n (\xi) + 2\xi S_n^\prime (\xi)\right)\right] = 0\eqno(13)$$
where $\xi = \frac{g^2}{\epsilon}$. Upon making the expansions
$$\frac{a_0(g)}{a_0^\prime (g)} = \frac{g + a_{1,0} g^3 + a_{2,0}g^5 +\ldots}{1 + 3a_{1,0} g^2 + 5a_{2,0}g^4 +\ldots} = g + \alpha_3 g^3 + \alpha_5 g^5 + \ldots\eqno(14)$$
and
$$\beta (g) = \sum_{k=1}^{\infty} B_{2k+1} g^{2k+1}\eqno(15)$$
then by looking at contributions to eq. (13) in ascending powers of $g$,
$$\left(1 - \xi B_3\right) S_0^\prime - \frac{1}{2} B_3 S_0 = 0\eqno(16)$$
$$\frac{1}{\xi} S_1 + \left(- \frac{1}{\xi} + B_3\right)\left( 3S_1 + 2\xi S_1^\prime\right) + \left( - \frac{\alpha_3}{\xi} + B_5\right)\left(S_0 + 2\xi S_0^\prime\right) = 0\eqno(17)$$
etc., whose solutions subject to eq. (7) are
$$S_0 (\xi) = w^{-1/2}\eqno(18)$$
$$S_1(\xi) = \frac{1}{w^{1/2}(1-w)}\left[\left(\frac{\alpha_3}{2} - \frac{B_5}{2B_3}\right) \left(-1 + w^{-1}\right) - \frac{B_5}{2B_3} \ln w\right],\eqno(19)$$
where $w = 1 - B_3\xi$. It is apparent that in the limit $\epsilon \rightarrow 0 (w \rightarrow \infty)$,
$$S_0 \rightarrow 0\;\;\;\;\;S_1 \rightarrow 0\eqno(20)$$
which is consistent with $S_k \rightarrow 0 (k \geq 2)$. This indicates that  the bare coupling $g_B$ vanishes as $\epsilon$ vanishes, which is not what might be expected from eq. (1). One can also derive this result from eq. (12), as upon multiply eq. (12) by $\epsilon^{-k+1}$ and summing over $k$ we arrive at 
$$\epsilon \left( g_B - \frac{a_0(g)}{a_0^\prime (g)} \frac{\partial}{\partial g} g_B\right) + \beta(g) \frac{\partial g_B}{\partial g} = 0\eqno(21)$$
so that
$$g_B = \mu^\epsilon \exp \left(- \int_K^g dx\frac{\epsilon}{\left(\beta (x) - \epsilon \frac{a_0(x)}{a_0^\prime (x)}\right)}\right)\eqno(22)$$
where $K$ is a cut off. Consistency with eqs. (18) and (19) is achieved by having
$$g_B = \mu^\epsilon \exp \left(- \int_0^g dx\left[
\frac{\epsilon}{\beta(x) - \epsilon \frac{a_0(x)}{a_0^\prime (x)}} + \frac{1}{x} \right]\right) .\eqno(23)$$
In the limit $\epsilon \rightarrow 0$, 
$$g_B \rightarrow g \exp \left(-\lim_{\delta \rightarrow 0^+} \int_\delta^g \frac{dx}{x} \right) = 0\eqno(24)$$
which is consistent with eq. (20).

The function $a_0(g)$ characterizes the choice of renormalization scheme; minimal subtraction uses a renormalized coupling $\widetilde{g}$ where
$$\widetilde{g} = a_0 (g).\eqno(25)$$
By eq. (10),
$$\mu \frac{d\widetilde{g}}{d\mu} = \widetilde{\beta} (\widetilde{g})
- \epsilon \widetilde{g} = \left(\frac{d\widetilde{g}}{dg}\right)\left(\mu \frac{dg}{d\mu}\right) = a_0^\prime (g) \left(\beta (g) - \epsilon \frac{a_0(g)}{a_0^\prime (g)}\right)\eqno(26)$$
so that
$$\widetilde{\beta}(\widetilde{g}) = a_0^\prime (g)\beta(g).
\eqno(27)$$
\noindent
{\bf\large(II)}  The ``method of characteristics'' can also be used to analyze the renormalization group equation, in particular eq. (8) [14,9,12,15-17]. We shall examine a simple example of how this technique can be used to extract information from a first order partial differential equation.

Suppose $A_0(x,y)$ is a solution to the equation
$$f(x,y) \frac{\partial A(x,y)}{\partial x} + g(x,y) \frac{\partial A(x,y)}{\partial y} + h (x,y) A(x,y) = 0\eqno(28)$$
where $f$, $g$ and $h$ are prescribed functions of $x$ and $y$. If now we have ``characteristic functions''  $\overline{x} (t)$ and $\overline{y}(t)$ defined by
$$\frac{d\overline{x}(t)}{dt} = f(\overline{x} (t), \overline{y} (t));\;\;\;\;\overline{x}(0) = x\eqno(29)$$
$$\frac{d\overline{y}(t)}{dt} = g(\overline{x} (t), \overline{y} (t));\;\;\;\;\overline{y}(0) = y\eqno(30)$$
then
$$A(\overline{x}(t), \overline{y}(t)) = A_0(\overline{x}(t), \overline{y}(t))\exp \int_0^t h(\overline{x}(t^\prime), \overline{y}(t^\prime))dt^\prime \eqno(31)$$
reduces to $A_0 (x,y)$ when $t = 0$ and $\frac{d}{dt} A(\overline{x}(t), \overline{y}(t)) = 0$.

For example, if $f = x$, $g = y^2$ and $h = 0$, a solution to eq. (28) is
$$A_0 (x,y) = xe^{1/y}\eqno(32)$$
while by eqs. (29) and (30),
$$\overline{x} (t) = xe^t\eqno(33)$$
$$\overline{y} (t) = \frac{y}{1-yt}.\eqno(34)$$
We indeed find that for all $t$
$$A_0 (\overline{x}(t), \overline{y}(t)) = \left(xe^t\right) \exp\left[\frac{1}{(y/(1-yt)}\right] = xe^{1/y}.\eqno(35)$$
However, if we knew only an approximate solution
$$A_0^{(1)} (x,y) = x\left(1 + \frac{1}{y}\right)\eqno(36)$$
then we might make the ansatz
$$A_0 (x,y) = x\sum_{n=0}^\infty \alpha_n \left(\frac{1}{y}\right)^n,\eqno(37)$$
with 
$$\alpha_0 = \alpha_1 = 1.\eqno(38)$$
The differential equation 
$$x \frac{\partial A_0}{\partial x} + y^2 \frac{\partial A_0}{\partial y} = 0\eqno(39)$$
shows that
$$\alpha_n = (n + 1) \alpha_{n+1} = \frac{1}{n!}\eqno(40)$$
and so
$$A_0 (x,y) = x\sum_{n=0}^\infty \frac{1}{n!} \left(\frac{1}{y}\right)^n = xe^{1/y}.\eqno(41)$$
Eq. (37) is analogous to eq. (5); in both cases we have knowledge of some part of a  solution to a first order partial differential equation, and the remaining portion of the solution can be determined by a recursion relation. In the equation
$$\left( \mu \frac{\partial}{\partial\mu} + \left(-\epsilon g + \beta (g)\right)\frac{\partial}{\partial g}\right) g_B = 0 \eqno(42)$$
we in general have only a partial knowledge of $\beta (g)$ itself; knowing $\beta (g)$ to $k$-loop order, as well using the fact that in the MS scheme $a_0(g) = g$, allows one to determine $S_0(\xi) \ldots S_k(\xi)$ in eq. (5) using eq. (42).

One could also use the characteristic function of eqs. (33) and (34) in conjunction with the approximate solution of eq. (36) to recover the exact solution of eq. (32). 
If we were to examine
$$A_0^{(1)} (\overline{x}(t), \overline{y}(t)) = \left(xe^t\right)
\left[ 1 + \frac{1}{(y/(1-yt))}\right]\eqno(43)$$
then it is apparent that at a particular value of $t$ (namely $t = \frac{1}{y}$) we have
$$A_0^{(1)} \left( \overline{x}\left(\frac{1}{y}\right), \overline{y}\left(\frac{1}{y}\right)\right) = A_0(x,y).\eqno(44)$$
(Furthermore, the ansatz of eq. (37) yields
$$A_0(\overline{x}(t),\overline{y}(t)) = xe^t \sum_{n=0}^\infty
\alpha_n\left(
\frac{1-yt}{y}\right)
\eqno(45)$$
which is independent of $t$ only if eq. (40) is satisfied, in which case we also recover eq. (32).)

An analogous treatment of eq. (42) using the characteristic functions $\overline{\mu}(t)$ and $\overline{g}(t)$ can also be performed. We first define
$$\frac{d\overline{\mu}(t)}{dt} = \overline{\mu}(t)\;\;\;\;\;(\overline{\mu}(0) = \mu)\eqno(46)$$
$$ \frac{d\overline{g}(t)}{dt} = -\epsilon\overline{g}(t)+ \beta(\overline{g}(t))\;\;\;(\overline{g}(0) = g).\eqno(47)$$
Equations (46) and (47) are solved iteratively using
an approach outlined in [15,12,17,18]. A rescaling
$$\epsilon \rightarrow \epsilon\lambda \;\;\;\;\;t \rightarrow t/\lambda \;\;\;\;\; \overline{g}(t) \rightarrow \overline{g}(t)\,\sqrt{\lambda}\eqno(48)$$
followed by an expansion
$$\overline{g}(t) = \overline{g}_0(t) + \overline{g}_1(t)\lambda + \overline{g}_2(t)\lambda^2 + \ldots\eqno(49)$$
with
$$\overline{g}_n(0) = g\delta_{n0}\eqno(50)$$
leads to
$$\overline{\mu}(t) = \mu e^{t/\lambda}\eqno(51)$$
as well as
$$\overline{g}_0(t) = g^2\left[
\frac{B_3}{\epsilon}\left(1 - 2e^{2\epsilon t}\right) + e^{2\epsilon t}\right]^{-1}\eqno(52)$$
$$\overline{g}_1(t) = \frac{B_5e^{2\epsilon t}g^5}
{2\epsilon\left[\frac{B_3g^2}{\epsilon} + \left(1 - \frac{B_3g^2}{\epsilon}\right)e^{2\epsilon t}\right]^{3/2}}
\left\lbrace \frac{1 - \frac{B_3}{\epsilon}g^2}{\left(\frac{B_3g^2}{\epsilon}\right)^2}\right.\eqno(53)$$
$$\left. \ln\left(\frac{\frac{B_3g^2}{\epsilon} + \left(1 - \frac{B_3g^2}{\epsilon}\right)e^{2\epsilon t}}{e^{2\epsilon t}}\right) + \frac{1}{\left(\frac{B_3}{\epsilon}g^2\right)}
\left(1 - \frac{1}{e^{2\epsilon t}}\right)\right\rbrace\nonumber$$
etc.
Together, eqs. (1), (3), (48) and (49) lead to 
$$g_B(\overline{\mu}(t), \overline{g}(t)) = \overline{\mu}^{\epsilon\lambda}
\left\lbrace\lambda^{1/2}\left[\overline{g}_0 + \frac{a_{11}\overline{g}_0^3}{\epsilon} + \frac{a_{22}\overline{g}_0^5}{\epsilon^2} + \ldots \right]\right.\eqno(54)$$
$$+ \lambda^{3/2}\left[\left(\overline{g}_1 + \frac{3a_{11}\overline{g}_0^2\,\overline{g}_1}{\epsilon} + \frac{5a_{22}\overline{g}_0^4\,\overline{g}_1}{\epsilon^2} + \ldots\right)\right.\nonumber$$
$$\left. \left. + \left(\frac{a_{21}\overline{g}_0^5}{\epsilon} + 
\frac{a_{31}\overline{g}_0^7}{\epsilon^2} + \ldots\right)\right] + O(\lambda^{5/2})\right\rbrace .\nonumber$$
In the limit $t \rightarrow \infty$ we find from eqs. (52) and (53) that
$$\overline{g}_0 \rightarrow g\left( 1 - \frac{B_3g^2}{\epsilon}\right)e^{-\epsilon t}\eqno(55)$$
$$\overline{g}_1 \rightarrow 
 \frac{B_5g^5}{2\epsilon} 
 \left(1- \frac{B_3g^2}{\epsilon}\right)^{-3/2}
 \left\lbrace 
 \frac{1 -\frac{B_3g^2}{\epsilon}}{\left(\frac{B_3g^2}{\epsilon}\right)^2} 
 \ln \left(1 - \frac{B_3g^2}{\epsilon}\right) + \frac{1}{\left(\frac{B_3g^2}{\epsilon}\right)}\right\rbrace e^{-et}\eqno(56)$$
  while
$$\overline{\mu}^{\epsilon\lambda} \rightarrow \mu^{\epsilon\lambda}e^{\epsilon t}.\eqno(57)$$
From eqs. (54 - 57) we find that as $t \rightarrow \infty$
$$g_B = \mu^{\epsilon\lambda} \left\lbrace \lambda^{1/2} \left[ \frac{g}{\sqrt{1 - \frac{B_3g^2}{\epsilon}}}\right] + \lambda^{3/2} \left[ \frac{B_5g^5}{2\epsilon} \frac{1}{\left(1- \frac{B_3g^2}{\epsilon}\right)^{3/2}}
\right.\right. \nonumber$$
$$\left.\left.
\times
\left( \frac{1-\frac{B_3g^2}{\epsilon}}{\left(\frac{B_3g^2}{\epsilon}\right)^2}
\ln \left(1 - \frac{B_3g^2}{\epsilon}\right) + \frac{1}{\left(\frac{B_3g^2}{\epsilon} \right)}\right)\right]\right\rbrace ,\eqno(58)$$
which when $\lambda = 1$ reproduces what was found in eqs. (5), (18) and (19). The method of characteristics should reproduce all of the $S_n(\xi)$ appearing in eq. (5) as $t \rightarrow \infty$. This is analogous to eq. (43) reducing to eq. (44) when $t = \frac{1}{y}$.
\\
\noindent
{\bf\large(III)} Having illustrated how the renormalization group equation works by considering the relationship between the bare and renormalized couplings when using dimensional regularization, we now will examine how it can be employed in conjunction with the perturbative calculation of a physical quantity [19-20].

In particular, if we consider $R(s)$, the ratio of the total cross section $\sigma$ ($e^+e^- \rightarrow$ (hadrons)) to the total cross section $\sigma(e^+e^- \rightarrow \mu^+\mu^-)$, we find that a perturbative calculation of $R(s)$ yields
$$R(S) = 3 \sum_f Q_f^2 S\left[x(\mu^2), \ln\left(\frac{\mu^2}{s}\right)\right]\eqno(59)$$
where $s$ is the centre of mass energy, $\mu^2$ is the renormalization scale and $x(\mu^2)$ is the coupling $\alpha_s(\mu^2)/\pi$. As $\mu^2$ is unphysical, we have the renormalization group equation
$$\mu^2 \frac{dR(s)}{d\mu^2} = 0 = \left(\mu^2 \frac{\partial}{\partial\mu^2} + \beta(x) \frac{\partial}{\partial x}\right) R(s)\eqno(60)$$
where
$$\beta (x(\mu^2)) = \mu^2 \frac{dx(\mu^2)}{d\mu^2} .\eqno(61)$$
Perturbation theory results in the expansions
$$\beta (x) = -x^2 \sum_{k=0}^\infty \beta_k x^k\eqno(62)$$
$$S[x,L] = 1 + \sum_{n=1}^\infty \sum_{m=0}^{n-1} T_{n,m} x^n L^m\eqno(63)$$
where $x = x(\mu^2)$ and $L = \ln\left(\frac{\mu^2}{s}\right)$. Reorganizing the sum in eq. (63) so that in analogy with eq. (5)
$$S[x,L] = 1 + \sum_{n=1}^\infty x^nS_n(xL)\eqno(64)$$
where
$$S_n(\xi) = \sum_{k=0}^\infty T_{n+k,k}\xi^k\eqno(65)$$
leads to
$$\left(1 - \beta_0\xi\right)\frac{dS_k}{d\xi} - k\beta_0 S_k = \left(1 - \delta_{k,1}\right)\sum_{\ell = 1}^{k-1} \beta_{\ell} \left(\xi \frac{d}{d\xi} - \ell + k\right) S_{k - \ell}\;\;\;\; \left( S_n(0) = T_{n,0}\right)\eqno(66)$$
upon substitution of eq. (64) into eq. (60). Solving these equations sequentially we obtain [20]
$$S_1 (\xi) = w^{-1}\;\;\;\;\;(w \equiv 1 - \beta_0\xi)\eqno(67)$$
$$S_2 (\xi) = \left[ T_{2,0} - \frac{\beta_1}{\beta_0} \ln w\right]w^{-2}\eqno(68)$$
$$S_3(\xi) = \left( \frac{\beta_1^2}{\beta_0^2} - \frac{\beta_2}{\beta_0}\right) w^{-2} + \left( T_{3,0}- \left(\frac{\beta_1^2}{\beta_0^2} - \frac{\beta_2}{\beta_0}\right) - \frac{\beta_1}{\beta_0} 
\left(2T_{2,0} + \frac{\beta_1}{\beta_0}\right)\ln w + \frac{\beta_1^2}{\beta_0^2} \ln^2 w\right)w^{-3}\eqno(69)$$
and
$$S_4(\xi) =- \frac{1}{2} \left[ \frac{\beta_1}{\beta_0}  \left(\frac{\beta_1^2}{\beta_0^2}- 2 \frac{\beta_2}{\beta_0}\right) +  \frac{\beta_3}{\beta_0}\right] w^{-2}+  
\left(2T_{2,0} + \frac{\beta_1}{\beta_0}\right)\left( \frac{\beta_1^2}{\beta_0^2} - \frac{\beta_2}{\beta_0}\right) w^{-3}\nonumber$$
$$ + 2 \left( \frac{\beta_1}{\beta_0}\right)\left(\frac{\beta_2}{\beta_0} - \frac{\beta_1^2}{\beta_0^2}\right) w^{-3} \ln w + \left[ T_{4,0} +  \frac{\beta_3}{2\beta_0}- \frac{1}{2} \frac{\beta_1^3}{\beta_0^3}\right.\eqno(70)$$
$$\left. -2T_{2,0} \left(\frac{\beta_1^2}{\beta_0^2}- \frac{\beta_2}{\beta_0}\right)\right]w^{-4} +  \frac{\beta_1}{\beta_0} 
\left[2   \frac{\beta_1^2}{\beta_0^2} - 3 \frac{\beta_3}{\beta_0} - 2T_{2,0} \frac{\beta_1}{\beta_0}-3T_{3,0}\right]w^{-4} \ln w\nonumber$$
$$ + \frac{\beta_1}{\beta_0} 
\left[\frac{5\beta_1^2}{2\beta_0^2} + 3T_{2,0} \frac{\beta_1}{\beta_0}\right] w^{-4}\ln^2w -  \frac{\beta_1^3}{\beta_0^3}w^{-4}\ln^3 w.\nonumber$$
$S_k(\xi)$ for $k > 4$ could be determined if $T_{k,0}$ and $\beta_{k-1}$ were known. $S_1$ is the ``leading log'' ($LL$) sum; $S_2$ is the ``next to leading log'' ($NLL$) sum etc.

The approximation to $R(s)$ given by 
$$S[x(\mu), L(\mu)] = 1 + \sum_{n=1}^4 x^n(\mu) S_n(x(\mu)L(\mu))\eqno(71)$$
where $x(\mu)$ is a solution to
$$\mu^2 \frac{dx(\mu)}{d\mu^2} = -x^2(\mu) \sum_{k=0}^3 \beta_k x^k(\mu)\eqno(72)$$
is virtually independent of $\mu$; using the purely perturbative result
$$S[x(\mu), L(\mu)] = 1 + \sum_{n=1}^4\sum_{m=0}^3 T_{n,m}x^n(\mu) L^m(\mu)\eqno(73)$$
has a pronounced dependence on $\mu$. 
This is not surprising, as eq. (72) is a solution of the renormalization group equation, with $\beta(x)$ truncated at four-loop order. This serves to demonstrate that the renormalization group improved expression for $R(s)$ given by eq. (71) has enhanced predictive power over the purely perturbative result of eq. (73), as physical results should be independent of the unphysical parameter $\mu^2$.

One can also employ the method of characteristics to recover the result of eq.(71) [17] by following the techniques used in deriving eq. (54) when the relationship between $g_B$ and $g$ was discussed. In association with eq. (60) we define characteristic functions $\overline{\mu}^2(t)$ and $\overline{x}(t)$ satisfying
$$\frac{d}{dt} \overline{\mu}^2(t) = \overline{\mu}^2(t)\;\;\;\;\;\left(\overline{\mu}^2(0) = \mu^2\right)\eqno(74)$$
$$\frac{d}{dt} \overline{x}(t) = \beta(\overline{x}(t))\;\;\;\;\;\left(\overline{x}(0) = x(\mu^2)\right).\eqno(75)$$
It is apparent from eqs. (61) and (75) that $x(\mu^2)$ and $\overline{x}(t)$, the running and characteristic functions, satisfy the same equations. They are however distinct functions with the running coupling serving as a boundary condition to the characteristic function.

We begin by a rescaling
$$t \rightarrow t/\lambda \eqno(76)$$
$$\overline{x} \rightarrow \overline{x}\lambda\eqno(77)$$
where $\lambda$ is a book keeping parameter, so that
$$\lambda \frac{d\overline{\mu}(t)}{dt} = \overline{\mu}^2(t)\eqno(78)$$

$$\lambda^2 \frac{d\overline{x}(t)}{dt} = -\lambda^2\overline{x}^2(t)\sum_{n=0}^\infty \overline{x}^n(t)\lambda^n\beta_n\eqno(79)$$
and
$$S[\overline{x}, \overline{L}] = 1 + \sum_{n=1}^\infty \sum_{m=0}^{n-1} T_{n,m}\lambda^n\overline{x}^{\,n}\overline{L}^{\,m}\eqno(80)$$
where $\overline{L} \equiv \ln\left(\overline{\mu}^2/S\right)$. Perturbatively expanding $\overline{x}(t)$
$$\overline{x}(t) = \sum_{n=0}^\infty \lambda^n\overline{x}_n(t)\;\;\;\;
\left(\overline{x}_n(0) = x(\mu^2)\delta_{n0}\right)\eqno(81)$$
means that eq. (79) can be satisfied order-by-order in $\lambda$ provided that
$$\overline{x}_0(t) = x\overline{w}^{-1}\;\;\left(x \equiv x(\mu^2),\;\;\;\;\overline{w} \equiv 1 + \beta_0 x(\mu^2)t\right)\eqno(82)$$
$$\overline{x}_1(t) = -\frac{\beta_1}{\beta_0} \left(\frac{x}{\overline{w}}\right)^2 \ln \overline{w}\eqno(83)$$
$$\overline{x}_2(t) = \left(\frac{x}{\overline{w}}\right)^3\left[
\left( \frac{\beta_1^2}{\beta_0^2} - \frac{\beta_2}{\beta_0}\right)\beta_0 xt - \frac{\beta_1^2}{\beta_0^2} \ln \overline{w} + \frac{\beta_1^2}{\beta_0^2} \ln^2 \overline{w}\right]\eqno(84)$$
$$\overline{x}_3(t) = \left(\frac{x}{\overline{w}}\right)^4\left[
\left( \frac{-\beta_1^3}{2\beta_0^3} + \frac{\beta_1\beta_2}{\beta_0^2} - \frac{\beta_3}{2\beta_0}\right)\overline{w}^2 + \left(\frac{\beta_1^3}{\beta_0^3} - \frac{\beta_1\beta_2}{\beta_0^2}\right)\overline{w}(1 - 2\ln\overline{w})\right.\nonumber$$
$$+ \left(-\frac{\beta_1^3}{2\beta_0^3} + \frac{\beta_3}{2\beta_0}\right) + \left(\frac{2\beta_1^3}{\beta_0^3}- \frac{3\beta_1\beta_2}{\beta_0^2}\right)\ln\overline{w}\nonumber$$
$$\left. + \frac{5\beta_1^3}{2\beta_0^3} \ln^2 \overline{w} - \frac{\beta_1^3}{\beta_0^3}\ln^3\overline{w}\right] .\eqno(85)$$
Together, eqs. (80) and (81) lead to
$$S[\overline{x}(t), \overline{L} (t)] = 1 + \lambda \left[ T_{1,0} \overline{x}_0\right] + \lambda^2\left[T_{1,0}\overline{x}_1 + \left(T_{2,0} + T_{2,1} \overline{L}\right)\overline{x}_0^2\right]\nonumber$$
$$+ \lambda^3\left[T_{1,0}\overline{x}_2 + \left(T_{2,0} + T_{2,1}\overline{L}\right)\left(2\overline{x}_0\overline{x}_1\right)\right.\nonumber$$
$$\left. + \left(T_{3,0} + T_{3,1} \overline{L} + T_{3,2} \overline{L}^2\right)\overline{x}_0^3\right]\nonumber$$
$$+ \lambda^4\left[ T_{1,0} \overline{x}_3 + \left(T_{2,0} + T_{2,1} \overline{L}\right)\left(\overline{x}_1^2 + 2\overline{x}_0\overline{x}_2\right)\right.\eqno(86)$$
$$+ \left(T_{3,0} + T_{3,1} \overline{L} + T_{3,2}\overline{L}^2\right)\left(3\overline{x}_0\overline{x}_1\right)\nonumber$$
$$\left. +\left(T_{4,0} + T_{4,1} \overline{L} + T_{4,2} \overline{L}^2 + T_{4,3} \overline{L}^3\right)\overline{x}_0^4\right] + \ldots\nonumber$$
As was shown in the discussion leading up to eq. (31), eq. (86) is independent of $t$, provided one were to sum all contributions to $S[\overline{x}(t), \overline{L}(t)]$. This independence of $t$ can be exploited by writing $t$ as $t = \lambda \ln k$ (so that by eq. (78) $\overline{\mu}^2(t) = k\mu^2$) and then expanding eq. (86) in powers of $k$; the resulting expression reduces to eq. (63) provided [17]
$$T_{2,1} = \beta_0 T_{1,0}\nonumber$$
$$T_{3,1} = 2T_{2,0}\beta_0 + \beta_1 T_{1,0}\nonumber$$
$$T_{3,2} = \beta_0^2 T_{1,0}\nonumber$$
$$T_{4,1} = 3\beta_0 T_{3,0} + 2\beta_1 T_{1,0} + \beta_2T_{1,0}\eqno(87)$$
$$T_{4,2} = 3\beta_0^2 T_{2,0} + \frac{5}{2} \beta_0 \beta_1 T_{1,0}\nonumber$$
$$T_{4,3} = \beta_0^3 T_{1,0}\nonumber$$
etc.
These relationships are precisely what one obtains if eqs. (62) and (63) were substituted directly into eq. (60). Furthermore, eqs. (87) and (65) can be shown to lead to eqs. (67-70). A more direct way of recovering eq. (71) is to choose $t = \lambda\ln \left(\frac{s}{\mu^2}\right)$ so that $\overline{\mu}^2 = s$, $\overline{L} = 0$ and $\overline{w} = w$; taking $\lambda = 1$ results in eq. (86) when truncated at $O\left(\lambda^4\right)$ reducing to eq. (71). This is much like what happened in eq. (44); in both cases a perturbative approximation to the solution of a linear partial differential equation becomes an exact solution when evaluated at the characteristic functions for a certain value of the characteristic parameter.

A number of other physical processes [19], among them the semi-leptonic decay rate of $b$ mesons, QCD contributions to vector and scalar correlation functions, the Higgs decay rate into two gluons and the perturbative static potential, can also be analyzed using the renormalization group. In each case, physical quantities when computed perturbatively, exhibit a dependence on the renormalization $\mu^2$; this dependence is virtually eliminated when a $k^{th}$ order perturbative calculation is improved by making use of the renormalization group equation to sum $N^kLL$ contributions.

\noindent
{\bf\large(IV)} Renormalization group improvement can also be employed in thermal field theory calculations [21]. In QCD, if there are $n_f$ quark flavours, the thermodynamic free energy for $T >> 0$ is [22,23,24]
$$F = \frac{-8\pi^2}{45} T^4 \left\lbrace
\left(1 + \frac{21}{32} n_f\right)+ \frac{-15}{4} \left(1 + \frac{5}{12} n_f\right) \frac{\alpha_s}{\pi} + 30 \left[\left(1 + \frac{n_f}{6}\right)\left(\frac{\alpha_s}{\pi}\right)\right]^{3/2}\right.\nonumber$$
$$+\left\lbrace 237.2 + 15.97n_f - .413n_f^2 + \frac{135}{2} \left(1 + \frac{n_f}{6}\right)\ln\left[\frac{\alpha_s}{\pi}\left(1 + \frac{n_f}{6}\right)\right]
\right.\eqno(88)$$
$$\left. - \frac{165}{8} \left(1 + \frac{5}{12} n_f\right)
\left(1 - \frac{2}{33} n_f\right)\ln \frac{\overline{\mu}}{2\pi T}\right\rbrace \left(\frac{\alpha_s}{\pi}\right)^2 + \left(1 + \frac{n_f}{6} \right)^{1/2}\nonumber$$
$$\left[ -799.2 - 21.96n_f - 1.926n_f^2 + \frac{495}{2} \left(1 + \frac{n_f}{6}\right)\left(1 - \frac{2}{33} n_f\right)\right.\nonumber$$
$$\left.\left. \ln \frac{\overline{\mu}}{2\pi T}\right] \left(\frac{\alpha_s}{\pi}\right)^{5/2} + O\left(\alpha_s^3 \ln \alpha_s\right)\right\rbrace\nonumber$$
where $\overline{\mu}^2$ is the ${\overline{MS}}$ renormalization scale, and $x(\overline{\mu}^2) = \frac{\alpha_s(\overline{\mu}^2)}{\pi}$ satisfies the two-loop renormalization group equation
$$\mu^2 \frac{dx(\mu^2)}{d\mu^2} = b_2 x^2 (\mu^2) + b_3x^3(\mu^2).\eqno(89)$$
Changes in $\overline{\mu}^2$ in eq. (88) (with corresponding changes in $\alpha_s$ dictated by eq. (89)) lead to large variations in $F$. To overcome this deficiency in the perturbative calculation of $F$, we look for a solution to the renormalization group equation
$$\left(\overline{\mu}^2 \frac{\partial}{\partial\overline{\mu}^2} + \beta(x) \frac{\partial}{\partial x}\right) \left(\frac{F}{F_0}\right) = 0\eqno(90)$$
that is of the form
$$F/F_0 = 1 + \sum_{n=0}^\infty \left(
R_n(\xi) x^{n+1} + S_n(\xi)x^{n+3/2} + T_n(\xi)x^{n+2} \ln x\right)\eqno(91)$$
where $F_0$ is the ideal-gas value of $F$, $\xi = x L = x \ln \left(\frac{\overline{\mu}^2}{(2\pi T)^2}\right)$.

The general form of the functions $R_n(\xi)$, $S_n(\xi)$ and $T_n(\xi)$ is
$$R_n(\xi) = \sum_{m=0}^\infty
A_{n+m,m}\xi^n,\;\;\; 
S_n(\xi) = \sum_{m=0}^\infty
B_{n+m,m}\xi^m,\;\;\;
T_n(\xi) = \sum_{m=0}^\infty
C_{n+m,m}\xi^m\eqno(92)$$
with the renormalization group equation (90) allowing us to determine 
$A_{n+m,m}$, $B_{n+m,m}$ and $C_{n+m,m}$ ($m > 0$) in terms of 
$A_{n,0}$, $B_{n,0}$, $C_{n,0}$ and the $\beta$-function coefficients. From eqs. (89-91) it follows that
$$0 = \sum_{n=0}^\infty \left\lbrace \left[ R_n^\prime x^{n+2} + \left(b_2 x^2 + \ldots\right)\left(\xi R_n^\prime + (n+1)R_n + xT_n\right)x^n\right]\right.\eqno(93)$$
$$+ \left[ S_n^\prime x^{n+5/2} + \left(b_2 x^2 + \ldots\right)\left(\xi S_n^\prime + \left(n+\frac{3}{2}\right)S_n \right)x^{n+1/2}\right]\nonumber$$
$$\left. + \left[ T_n^\prime x^{n+3} + \left(b_2 x^2 + \ldots\right)\left(\xi T_n^\prime + \left(n+2\right)T_n \right)x^{n+1}\ln x\right]\right\rbrace .\nonumber$$
It is possible to read off $A_{0,0}$, $A_{1,0}$, $B_{0,0}$, $B_{1,0}$ and $T_{0,0}$ from eq. (88); these values combined with the requirement that eq. (93) be satisfied at each order in $x$ results in
$$R_0(\xi) = A_{0,0}\, w^{-1}, \;\;\; S_0(\xi) = B_{0,0}\, w^{-3/2}, \;\;\;
T_0(\xi) = C_{0,0}\, w^{-2}\eqno(94)$$
and
$$R_1(\xi) = w^{-2}\left[A_{1,0} - \left(\frac{b_3}{b_2} \,A_{0,0} + C_{0,0}\right)\ln w  \right],\;\;
S_1(\xi) = w^{-5/2}\left[B_{1,0} - \frac{3}{2}\,\frac{b_3}{b_2} \,B_{0,0} \ln w  \right]\eqno(95)$$
where $w = 1 + b_2\xi$. With these functions, the approximation to $F/F_0$ is given by
$$F/F_0 = 1 + \sum_{n=0}^1 \left( R_n(\xi)x^{n+1} + S_n(\xi)x^{n + 3/2}\right) + T_0(\xi)x^2 \ln x\eqno(96)$$
is virtually independent of $\overline{\mu}^2$ and has a value which coincides with the results of lattice calculations of this quantity. The effect of absorbing changes if $\overline{\mu}^2$ by altering $A_{0,0}$, $B_{0,0}$ and $C_{0,0}$ rather than having it simply change $\alpha_s$ according to eq. (89) is considered in [21], although this is contrary to the renormalization group approach.

The method of characteristics can also be used to arrive at this result [12]. As has been done in previous examples, characteristic functions $\overline{\mu}^2(t)$ and 
$\overline{x}(t)$ are introduced
$$\frac{d\overline{\mu}^2(t)}{dt} = \overline{\mu}^2(t)\;\;\;\;\left(\overline{\mu}^2(0) = \mu^2\right)\eqno(97)$$
$$\frac{d\overline{x}(t)}{dt} = b_2\overline{x}^2(t) + b_3 \overline{x}^3(t) 
+ \ldots (\overline{x} (0) = x(\mu^2))\eqno(98)$$
and then a rescaling
$$t \rightarrow t/\lambda ,\;\;\;\; \overline{x} (t) \rightarrow \lambda \overline{x} (t) \eqno(99)$$
is performed, followed by the perturbative expansion
$$\overline{x} (t) = \sum_{n=0}^\infty \lambda^n \overline{x}_n (t) \;\;\;\;
\left( \overline{x}_n (0) = x(\mu^2)\delta_{n,0}\right).\eqno(100)$$
Solving for $\overline{x}_n(t)$ iteratively and then substitution of the results into the perturbative result
$$F/F_0 = 1 + \left(A_{0,0} x + A_{1,0} x^2 + A_{1,1} x^2 L\right.\nonumber$$
$$\left. + B_{0,0} x^{3/2} + B_{1,0} x^{5/2} + B_{1,1}x^{5/2}L + C_{0,0} x^2 \ln x\right)\eqno(101)$$
(which is eq. (88)) reduces to what was obtained in eq. (96)
when $t = \lambda \ln \left(\frac{(2\pi T)^2}{\overline{\mu}^2}\right)$.

\noindent
{\bf\large(V)} When the renormalization group function $\beta$ is known exactly, information can be extracted about the effective action. This is the converse of the analysis of ref. [25] where computation of the two-loop effective action in both scalar and spinor quantum electrodynmaics in the presence of a self-dual background field strength was used to determine the two-loop $\beta$-function.

It has been shown that the effective action $L$ and the $\beta$-function are related by the equation
$$L = \frac{-1}{4} \frac{g^2}{\overline{g}^2(t,g)} \Phi\eqno(102)$$
where $\Phi$ is related to the constant background field strength $F_{\mu\nu}^a$ by
$$\Phi = F_{\mu\nu}^a F^{a\mu\nu}\eqno(103)$$
and
$$t = \frac{1}{4} \ln \left(\frac{g^2\Phi}{\mu^4}\right)\eqno(104)$$
$$ = \int_g^{\overline{g}(t,g)} \frac{dx}{\beta(x)} .\eqno(105)$$
Eq. (102) follows from the trace anomaly condition
$$\left\langle \theta^\mu_{\;\;\mu}
\right\rangle = \frac{\beta(\overline{g})}{2\overline{g}} \left(\frac{g}{\overline{g}}\right)^2 \Phi\eqno(106)$$
and the definition of the expectation value of the energy-momentum tensor

$$\left\langle \theta^{\mu\nu}
\right\rangle = -g^{\mu\nu} L + 2 \frac{\delta L}{\delta g_{\mu\nu}} .
\eqno(107)$$
$L$ in eq. (102) satisfies the renormalization group equation
$$\left(\mu \frac{\partial}{\partial\mu} + \beta (g) \frac{\partial}{\partial g} + \gamma (g) F_{\alpha\beta}^a \frac{\partial}{\partial F_{\alpha\beta}^a}\right) L = 0\eqno(108)$$
provided
$$\beta(g) =-g\gamma(g).\eqno(109)$$
Eq. (109) follows from the fact that [26,27] $gA^a_\mu$ is not renormalized in order to preserve gauge invariance in the background field.

We note that from eqs. (104) and (105),
$$\frac{\partial \overline{g}(t,g)}{\partial t} = \beta(\overline{g} (t,g)) = 
\frac{\partial \overline{g}(t,g)}{\partial t} \beta(g)\eqno(110)$$
and
$$\frac{\partial t}{\partial \mu} = -1.\eqno(111)$$

In the $SU(3) N = 1$ super Yang-Mills theory, the $\beta$-function in a supersymmetric renormalization scheme is given by [28,29] 
$$\beta (g) = \frac{-9g^3/(4\pi)^2}{1-6g^2/(4\pi)^2} .\eqno(112)$$
(It is not immediately apparent how this renormalization scheme is related to minimal subtraction [30].)  If $\overline{y} = -(4\pi)^2/6\overline{g}^2(t,g)$ and 
$y = -(4\pi)^2/6g^2$, then by eq. (105)
$$\overline{y} e^{\overline{y}} = e^{-3t}ye^y,\eqno(113)$$
so that if $W(\eta)e^{W(\eta)} = \eta$ defines the Lambert $W$ function [31], then eq. (102) becomes
$$L = \frac{-1}{4y} W\left[\left(\frac{g^2\Phi}{\mu^4}\right)^{-3/4}ye^y\right]\Phi .\eqno(114)$$

If we consider an $SU(3)$ $N = 2$ super Yang-Mills theory, then the $\beta$-function is an entirely one-loop quantity,
$$\beta (g) = - \frac{6}{(4\pi)^2} g^3 ;\eqno(115)$$
with this it follows from eqs. (102)and (105) that
$$L = \frac{-1}{4} \left[1 + \frac{3g^2}{(4\pi)^2} \ln \left( \frac{g^2\Phi}{\mu^4}\right)\right]\Phi .\eqno(116)$$

\noindent
{\bf\large(VI)} Instanton contributions to the effective action will now be examined; renormalization group considerations again show how a purely perturbative calculation can be improved. The one instanton contribution to the effective action in an $SU(2)$ gauge theory with $n_f$ flavours is of the form [32]
$$L_{eff} = K \int d\rho \;\rho^{-5+3n_f} \exp\left\lbrace \frac{-8\pi^2}{g^2(\mu)} S\right\rbrace\eqno(117)$$
where $S$ is a power series
$$S = 1 + \sum_{n=1}^{\infty} \sum_{m=0}^n T_{n,m}g^{2n}(\mu)\ln^m(\mu\rho).\eqno(118)$$
The integral over $\rho$, the ``size'' of the instanton, converges in the ultraviolet limit $(\rho \rightarrow 0)$ but diverges in the infrared limit $(\rho \rightarrow \infty)$ when the one-loop contributions $T_{1,0}$ and $T_{1,1}$ computed in [32] are included.

Resumming the two series in eq. (118) [18] so that
$$S = \sum_{k=0}^\infty g^{2k} (\mu) \left(\sum_{\ell = k}^\infty T_{\ell,\ell - k}\left(g^2(\mu)\ln (\mu\rho)\right)^{\ell - k}\right)\nonumber$$
$$\equiv \sum_{k = 0}^\infty g^{2k} (\mu) S_k \left(g^2(\mu)\ln(\mu\rho)\right)\eqno(119)$$
and applying the renormalization group equation
$$\left(\mu\frac{\partial}{\partial\mu} + \beta (g) \frac{\partial}{\partial g}\right) S = 0\eqno(120)$$
with the $\beta$-function being given by
$$\beta(g) = \sum_{\ell = 1}^\infty b_{2\ell + 1} g^{2\ell + 1}\eqno(121)$$
we find that
$$S_0 =1 + 2 b_3 g^2 (\mu) \ln (\mu\rho)\equiv w \eqno(122)$$
and for $k > 1$
$$\frac{dS_k}{dw} + \frac{k-1}{w} S_k = \frac{-1}{2b_3 w} \left[
\sum_{\ell = 0}^{k-1} b_{3 + 2(k - \ell)} \left( 2(\ell - 1) S_\ell 
+ 2 (w - 1) \frac{dS_\ell}{dw}\right)\right]\eqno(123)$$
with $S_k(w = 1) = T_{k,0}$. From eq. (123) it follows that
$$S_1 = T_{1,0} + \frac{b_5}{b_3}\ln |w|\eqno(124)$$
$$S_2 = \left(\frac{b_7}{b_3} - \frac{b_5^2}{b_3^2}\right) + 
\frac{T_{2,0} - \left(\frac{b_7}{b_3} - \frac{b_5^2}{b_3^2}\right) + 
 \frac{b_5^2}{b_3^2} \ln |w|}{w}.\eqno(125)$$
In general, from eq. (123) it follows that for $k \geq 2$,
$$S_k (w) = C_k + O\left(\frac{1}{w}\right)\eqno(126)$$
where $C_k$ is dependent only on the $\beta$-function coefficients. Consequently, by eq. (119),
$$\rho^{-5+3n_{f}} \exp \left( \frac{-8\pi^2}{g^2}S\right)\stackrel{\longrightarrow}{_{(\ln \rho \rightarrow \infty)}} \;\rho^{\frac{7}{3}\left(1 + n_f\right)} |\ln \rho |^{-8\pi^2 b_5/b_3}.\eqno(127)$$
We consequently see that even upon including the contribution of all terms in eq. (119) to $S$, the integral over $\rho$ still suffers from an infrared divergence as $\rho \rightarrow \infty$.

\noindent
{\bf\large(VII)}  It is also possible to use the renormalization group to improve a perturbative calculation of the effective potential [33-37]. This procedure has been applied in both scalar electrodynamics and the standard model at leading-log order 
[38-39] and beyond [40]. We will illustrate this procedure by considering leading-log corrections to the one-loop calculation of the effective potential in massless scalar electrodynamics.

In this model, the classical Lagrangian is given by
$$L = \frac{1}{2} \left(\partial_\mu \phi_1 - e A_\mu\phi_2\right)^2 + \frac{1}{2} \left(\partial_\mu \phi_2 + eA_\mu \phi_1\right)^2 - \frac{\lambda}{4!} \left(\phi_1^2 + \phi_2^2\right)^2.\eqno(128)$$
One-loop corrections to $V = \frac{\lambda}{4!} \left(\phi_1^2 + \phi_2^2\right)^2 \equiv \frac{\lambda\phi^4}{4!}$ in the Landau gauge result in [33]
$$V = \phi^4 \left[ \frac{\lambda}{4!} + \left(\frac{5\lambda^2}{1152\pi^2} + \frac{3e^4}{64\pi^2}\right)\left(\ln \frac{\phi^2}{\mu^2} + k\right) + O(\lambda^3 , e^6)\right]\eqno(129)$$
where $k$ is a constant whose value is fixed by a renormalization condition. If we chose this condition to be
$$\frac{d^4V}{d\phi^4} = \lambda\eqno(130)$$
when $\phi^2 = \mu^2$, then
$$k =-\frac{25}{6} .\eqno(131)$$
If now $<\phi>$ is the vacuum expectation value of $\phi$, then upon choosing 
$\mu^2 = <\phi>^2$, it follows from eq. (129) that if
$$\left. \frac{dV}{d\phi}\right|_{<\phi>} = 0\eqno(132)$$
then at the scale $\mu^2 =  \,<\phi>^2$
$$\lambda = \frac{33e^4}{8\pi^2} \eqno(133)$$
so that eq. (129) becomes
$$V = \frac{3e^4}{64\pi^2}\, \phi^4 \left[\ln \left(\frac{\phi^2}{<\phi>^2}\right) - \frac{1}{2}\right] + O (e^6).\eqno(134)$$
Upon identifying thescalar and vector masses with
$$m_\phi^2 = V^{\prime\prime} (\phi = \,<\phi>)\;\;\;\;\;\;\;\;m_A^2 = e^2 <\phi>^2\eqno(135)$$
respectively, we find from eq. (134) that [33]
$$\frac{m_\phi^2}{m_A^2} = \frac{3e^2}{8\pi^2}\eqno(136)$$
where in eq. (136) (as in eq. (134)) $e^2$ is evaluated at $\mu^2 = \,<\phi>^2$.

The general form of the effective potential is
$$V = \frac{\pi^2\phi^4}{6} S(\lambda , e^2, L)\eqno(137)$$
where $L = \ln \left(\frac{\phi^2}{\mu^2}\right)$. The renormalization group equation ${\mu} \,\frac{dV}{d\mu} = 0$ then gives rise to the equation
$$\left[(-2 + 2\gamma) \frac{\partial}{\partial L} + \beta_e \frac{\partial}{\partial e^2} + \beta_\lambda \frac{\partial}{\partial \lambda} + 4\gamma\right] S = 0\eqno(138)$$
with at  one-loop order [33]
$$\gamma = \frac{3e^2}{16\pi^2} ,\eqno(139)$$
$$\beta_e = \frac{e^4}{24\pi^2} ,\eqno(140)$$
$$\beta_\lambda = \frac{5\lambda^2}{24\pi^2} - \frac{3\lambda e^2}{4\pi^2} + \frac{9e^4}{4\pi^2} .\eqno(141)$$
$S$ in eq. (137) can now be expressed as a power series in $\lambda$, $e^2$ and $L$ with the leading-log  contribution  (which is fixed entirely by the one-loop calculation) being given by those terms in which the power $L$ is raised to being one less than the sum of the powers to which $e^2$ and $\lambda$ are raised. If $x = 4\pi^2e^2$, $y = 4\pi^2\lambda$ the Leading-Log contribution to $S$ is of the form
$$S_{LL} = \sum_{n=0}^\infty \left(R_{n,n-1} y^n L^{n-1} + \sum_{k=0}^\infty T_{n,k} x^ny^k L^{n+k-1}\right)\eqno(142)$$
Eq. (138) now becomes
$$\left[-2 \frac{\partial}{\partial L} +\left( \frac{5}{6} y^2 - 3xy + 9x^2\right) \frac{\partial}{\partial y} + \frac{x^2}{6} \frac{\partial}{\partial x} + 3x\right]S_{LL} (x, y, L) = 0.\eqno(143)$$

Together, eqs. (142) and (143) are satisfied at order,
$$y^p L^{p-1} \;\;{\rm{if}}\;\; -2 (p-1) R_{p,p-1} + \frac{5}{6} (p-1) R_{p-1,p-2} = 0;\eqno(144)$$
$$xy^p L \;\;{\rm{if}}\;\; -2pT_{1,p}  + \frac{5}{6} (p-1) T_{1,p-1} -3(p-1)R_{p,p-1} = 0; (p \geq 1)\eqno(145)$$
$$x^2y^p L^p \;\;{\rm{if}}\;\; -2(p+1)T_{2,p}  + \frac{5}{6} (p-1) T_{2,p-1} -3pT_{1,p}\nonumber$$
$$+ 9(p+1) R_{p+1,p} + \frac{19}{6} T_{1,p}  = 0\;\; (p \geq 1)\eqno(146)$$
$$x^ny^p L^{n+p-2} \;\;{\rm{if}}\;\; -2(p+n-1)  + \frac{5}{6} (p-1) T_{n,p-1} -3pT_{n-1,p}\nonumber$$
$$+ 9(p+1) T_{n-2,p+1} + \frac{n+17}{6} T_{n-1,p}  = 0\;\; (n \geq 3, p \geq 1)\eqno(147)$$
and
$$-2(n-1)T_{n,0} 
+ 9 T_{n-2,1} + \frac{n+17}{6} T_{n-1,0}  = 0\;\; (n \geq 3).\eqno(148)$$
From eq. (129), we have the values of $T_{0,1}$, $R_{1,0}$, $T_{2,0}$, $T_{0,2}$ as well as $T_{1,1} = 0$. These values are consistent with eqs.
 (144-148), and furthermore, if
$$S_0(yL) = \sum_{n=1}^\infty R_{n,n-1} (yL)^{n-1},\eqno(149)$$
$$S_j(yL) = \sum_{n=0}^\infty T_{j,n} (yL)^n \eqno(150)$$
we find that these recursion relations serve to fix the functions
$S_n(yL)$. It can be shown that in particular
$$S_0 (yL) = \frac{1}{1 - \frac{5}{12}yL} \equiv \frac{1}{w}\eqno(151)$$
$$S_1(w) = -\frac{9}{5} \left(\frac{w-1}{w}\right)^2\eqno(152)$$
$$S_2(w) = \frac{1}{20w^3} \left(20w^3 + 77w^2 - 34w + 27\right)\eqno(153)$$
and
$$S_3(w) = \frac{1}{240w^4} \left(580w^4 + 760w^3 - 323w^2 + 126w - 243\right).\eqno(154)$$
With
$$V^{LL} = \frac{\pi^2\phi^4}{6} \left[y S_0 (yL) + \sum_{n=1}^\infty x^n L^{n-1} S_n (yL)\right]+ K\phi^4\eqno(155)$$
the condition of eq. (130) leads to
$$\frac{6K}{\pi^2} = -\left(\frac{125}{72} y^2 + \frac{75}{4} x^2\right) - \left(\frac{2625y^3 + 1875y^4 + 625y^5}{1296}\right)\eqno(156)$$
$$+\frac{x\left(175y^2 + 250y^3 + 125y^4\right)}{48}\nonumber$$
$$-\frac{x^2\left(9450y + 10275y^2 + 5275y^3\right)}{432},\nonumber$$
upon including $S_1$ and $S_2$ in the sum of eq. (155). (At lowest order, this is consistent with eq. (131).) If at $\mu^2 = <\phi>^2$, where $L = 0$, the condition of eq. (132) leads to a generalization of eq. (133)

$$0 = \left[\left(\frac{1296y - 1980y^2 - 2625y^3 - 1875y^4 - 625y^5}{1944}\right)\right.\eqno(157)$$
$$+x\left(\frac{175y^2 + 250y^3 + 125y^4}{72}\right)\nonumber$$
$$\left. -x^2\left(\frac{7128 + 9450y + 10275y^2 + 5275y^3}{648}\right)\right].\nonumber$$

This defines a relationship between $x$ and $y$ when these running couplings are evaluated at $\mu^2 = <\phi>^2$. As eq. (157) is quadratic in $x$, there are two values of $y$ for each value of $x$, one large and the other small. If $y \leq .01$, then eq. (157) reduces to eq. (133).

We also find that if we use the potential of eq. (155) (with only $S_1$ and $S_2$ included) and $K$ given by eq. (156), then the analogue of eq. (136) is
$$\frac{m_\phi^2}{m_A^2} = \left( \frac{y}{2x} - \frac{27x}{4}\right) - \frac{1}{2592x} \left[1620y^2 + 2475y^3 + 1875y^4\right.\nonumber$$
$$+625y^5 - x\left(4455y^2 + 6750y^3 + 3375y^4\right)\eqno(158)$$
$$\left. + x^2 \left(26730y + 30825y^2 + 15825y^3\right)\right].\nonumber$$
In the region in which eq. (157) reduces to eq. (133), eq. (158) reduces to eq. (136).

The analysis that has just been used to examine the effective potential in massless scalar electrodynamics can also be applied in the standard model [38-39]. The dominant couplings that must be included are the top quark Yukawa coupling, the quartic scalar coupling and the strong gauge coupling; the effects of the $SU(2) \times U(1)$ gauge couplings are treated as being secondary. If the tree level mass of the scalar is taken to vanish, then upon including leading log contributions to the effective potential and using experimental values for the top quark Yukawa coupling and the strong gauge coupling, it is possible to show that the Higgs scalar would have a mass of about 216 GeV. Furthermore, the value of the quartic scalar coupling is considerably enhanced, leading to an increase in the scattering cross section $\sigma(W_L^+ W_L^- \rightarrow Z_L^0Z_L^0)$ by a factor of about 30 over conventional expectations. These results are quite stable when the accessible higher order effects are included [40].

\noindent
{\bf\large(VIII)} We now turn to an exact solution of the renormalization group equation, focusing on a massless $\lambda\phi_4^4$ theory first of all, then considering the effect of including a mass for this scalar field.

The renormalization group equation for the effective potential $V$ in this model is [33,15,16,41-43]
$$\mu\frac{dV(\mu,\lambda , \phi)}{d\mu} = \left(\mu \frac{\partial}{\partial\mu} + \beta(\lambda)\frac{\partial}{\partial\lambda} + \gamma(\lambda) \phi\frac{\partial}{\partial\phi}  \right)V(\mu,\lambda,\phi) = 0, \eqno(159)$$
where $\lambda = \lambda(\mu)$ and $\phi = \phi(\mu)$ satisfy
$$\mu\frac{d\lambda}{d\mu} = \beta(\lambda)\eqno(160)$$
and
$$\mu\frac{d\phi}{d\mu} = \gamma(\lambda)\phi .\eqno(161)$$
Upon taking
$$V(\mu,\lambda,\phi) = Y(\lambda , L)\phi^4\eqno(162)$$
where $L \equiv \ln \left(\frac{\lambda\phi}{\mu}\right)$, eq. (159) becomes
$$\left[ \frac{\partial}{\partial L} - \tilde{\beta} (\lambda) \frac{\partial}{\partial\lambda} - 4\tilde{\gamma}(\lambda)\right]Y(\lambda, L) - 0\eqno(163)$$
where
$$\tilde{\beta} = \beta/(1-\gamma -\beta/2\lambda), \;\;\; 
\tilde{\gamma} = \gamma/(1-\gamma -\beta/2\lambda).\eqno(164)$$

(The factor of $\lambda$ in the argument of $L$ arises if one uses dimensional regularization [43].) An auxiliary function $\overline{\lambda}(L,\lambda)$ is defined by
$$L = \int_\lambda^{\overline{\lambda}(L,\lambda)} \frac{dx}{\tilde{\beta}(x)}\eqno(165)$$
satisfies
$$\overline{\lambda} (L = 0, \lambda) = \lambda \eqno(166)$$
$$\frac{\partial\overline{\lambda}(L,\lambda)}{\partial L} = \tilde{\beta}(\overline{\lambda}(L,\lambda))\eqno(167)$$
$$\frac{\partial\overline{\lambda}(L,\lambda)}{\partial \lambda} = \frac{-\tilde{\beta}(\overline{\lambda}(L,\lambda))}{\tilde{\beta}(\lambda)}.
\eqno(168)$$

Often, eq. (163) is taken to imply [33,5,10,44]
$$Y(\lambda , L) = f(\overline{\lambda}(L,\lambda))\exp\left[4\int_0^L \tilde{\gamma} (\overline{\lambda}(x,\lambda))dx\right]; \eqno(169)$$
however, substitution of eq. (169) into the left side of eq. (163) results in
$$I = \left\lbrace f^\prime (\overline{\lambda}(L,\lambda))\left[\frac{\partial\overline{\lambda}(L,\lambda)}{\partial L} - \tilde{\beta} (\lambda)\frac{\partial\overline{\lambda}(L,\lambda)}{\partial\lambda}\right]\right.\eqno(170)$$
$$+ f(\overline{\lambda}(L,\lambda))\left[4\tilde{\gamma}(\overline{\lambda}(L,\lambda)) - 4\tilde{\gamma}(\lambda)\right.\nonumber$$
$$\left.\left.+ 4 \int_0^L \frac{\partial\tilde{\gamma}(\overline{\lambda}(x,\lambda))}{\partial\lambda} dx\right]\right\rbrace \exp\left[4 \int_0^L \tilde{\gamma}(\overline{\lambda}(x,L))) dx\right]\nonumber$$
which only vanishes if $L = 0$ (where eq. (163) is satisfied by eq. (169) because of eqs. (166-168)).

There is, however, a way of obtaining a formal solution to eq. (163) [12]. If
$$Y(\lambda , L) = \sum_{m=1}^\infty \sum_{n=0}^{m-1} a_{mn}\lambda^mL^n\eqno(171)$$
$$\equiv \sum_{m=0}^\infty A_m(\lambda)L^m\nonumber$$
then from eq. (163) we find that
$$A_{n+1} (\lambda) = \frac{1}{2(n+1)} \left[ \tilde{\beta} (\lambda) \frac{\partial}{\partial\lambda} + 4 \tilde{\gamma}(\lambda)\right]A_n(\lambda).\eqno(172)$$
If now
$$A_n (\lambda) = \exp \left[ - 4 \int_{\lambda_{0}}^\lambda dx \frac{\tilde{\gamma}(x)}{\tilde{\beta}(x)}\right]\tilde{A}_n(\lambda)\eqno(173)$$
and
$$\eta(\lambda) = 2 \int_{\lambda_{0}}^\lambda \frac{dx}{\tilde{\beta}(x)}\eqno(174)$$
then eq. (172) becomes
$$\tilde{A}_{n+1}(\eta) = \frac{1}{n+1} \frac{\partial}{\partial\eta} \tilde{A}_{n}(\eta).\eqno(175)$$
As a result of eqs. (171-175) we find that
$$Y(\lambda ,L) = \exp\left[-4 \int_{\lambda_{0}}^\lambda dx \frac{\tilde{\gamma}(x)}{\tilde{\beta}(x)}\right] \sum_{n=0}^\infty \tilde{A}_n(\lambda)L^n\nonumber$$
$$= \exp\left[-4 \int_{\lambda_{0}}^\lambda dx \frac{\tilde{\gamma}(x)}{\tilde{\beta}(x)}\right] \sum_{n=0}^\infty
\frac{L^n}{n!} \left(\frac{d}{d\eta}\right)^n 
 \tilde{A}_0(\lambda(\eta))\nonumber$$
$$= \exp\left[-4 \int_{\lambda_{0}}^\lambda dx \frac{\tilde{\gamma}(x)}{\tilde{\beta}(x)}\right] \tilde{A}_0(\lambda(\eta + L))\nonumber$$
$$= \exp\left[-4 \int_{\lambda(\eta+L)}^\lambda dx \frac{\tilde{\gamma}(x)}{\tilde{\beta}(x)}\right] A_0(\lambda(\eta + L)).\eqno(176)$$
Eqs. (169) and (176) are distinct.

We have succeeded in expressing the effective potential in terms of the function $A_0$; we can now find $A_0$ itself by imposing a second condition on the effective potential. This condition is provided by eq. (132). Together, eqs. (132) and (171) lead to 
$$\sum_{n=0}^\infty \left(4A_n(\lambda(\mu)) + 2(n+1)A_{n+1}(\lambda(\mu))\right)
\left(\ln\frac{\lambda<\phi>^2}{\mu^2}\right)^n <\phi>^3 = 0.\eqno(177)$$
When dealing with massless scalar electrodynamics, the analogue of this result at one-loop order leads to eq. (133) [33] which serves to relate the value of the quartic scalar coupling to the gauge coupling; this is extended to the leading log effective potential in eq. (157) above [38-39]. In the purely scalar $\lambda\phi^4_4$ model, the one-loop effective potential gives an inconsistent result [33]. A different interpretation of eq. (132) [45] is to take it to define a relationship between the function $A_0\left(\lambda\left(\mu^2 = \lambda(\mu^2)<\phi>^2\right)\right)$ and
$A_1\left(\lambda\left(\mu^2 = \lambda(\mu^2)<\phi>^2\right)\right)$, as if 
$\mu^2 = \lambda(\mu^2)<\phi>^2$ is the  value chosen for $\mu^2$, then all terms in the sum appearing in eq. (177) vanish except for the $n = 0$ term. We take the resulting equation
$$A_1\left(\lambda\left(\mu^2\right)\right) = -2
A_0\left(\lambda\left(\mu^2\right)\right)\eqno(178)$$
to hold for all $\mu^2$ as the actual value of $\lambda(\mu^2)$, for any value of $\mu^2$, is contingent upon the boundary condition imposed on the equation for running coupling and is hence arbitrary.

Together, eqs. (172) with $n = 0$ and (178) imply that
$$A^\prime_0 (\lambda) + \left(\frac{4}{\beta(\lambda)} - \frac{2}{\lambda}\right)A_0(\lambda) = 0\eqno(179)$$
and so
$$A_0 (\lambda) = K \exp \left[- \int_{\lambda_{0}}^\lambda \left(\frac{4}{\beta(x)} - \frac{2}{x}\right)dx\right]\eqno(180)$$
$$= K^\prime \lambda^{(2+4b_3/b_2^2)}\exp\left(\frac{4}{b_2\lambda}\right)
\exp \left\lbrace - \int_0^\lambda \left[\frac{4}{\beta(x)} - 4 \left(\frac{1}{b_2x^2} - \frac{b_3}{b_2^2x}\right)\right]dx\right\rbrace.\eqno(181)$$
(Here $K = A_0(\lambda_0)$ is a boundary value; the divergence appearing in eq. (180) as $\lambda_0 \rightarrow 0$ is absorbed into $K^\prime$ in eq. (181) with $\beta(\lambda) = b_2 \lambda^2 + b_3 \lambda^3 + \ldots$ .)

Substitution of eq. (180) into eq. (176) leads to 
$$V = A_0(\lambda)\exp\left(  \int_\lambda^{\lambda(\eta + L)} \left(\frac{4\gamma(x) - 4 + 2\beta(x)/x}{\beta(x)} \right)dx
\right)\phi^4 .\eqno(182)$$
By eqs. (164) and (174), eq. (182) collapses to
$$V = A_0 (\lambda)\exp -2 \left[\eta \left(\lambda (\eta + L)\right) - \eta\right]\phi^4\nonumber$$
$$= A_0 (\lambda) \exp (-2L)\phi^4 = \frac{A_0(\lambda)\mu^4}{\lambda^2}.\eqno(183)$$
This result is independent of $\phi$ (and is hence ``trivial'' [46]) and has non-analytic dependence on $\lambda$, but still satisfies eq. (159). Eq. (183) can also be shown to follow if one were to use
counterterm renormalization so that $L$ appearing in eq. (161) is given by $L = \ln \left(\frac{\phi^2}{\mu^2}\right)$ without any dependence on $\lambda$.

It appears that the only way to escape the conclusion that $V$ is independent of $\phi$ is to have eq. (177) satisfied not by eq. (178), but rather $<\phi> = 0$, in which case it is not possible to determine the function $A_0$, but there is no radiatively driven breaking to the symmetry $\phi \rightarrow -\phi$ in the original Lagrange density.

An alternate way of arriving at eq. (183) is to have eq. (177) satisfied order by order in $\ln \left(\frac{\lambda<\phi>^2}{\mu^2}\right)$, so that 
$$A_{n+1}(\lambda (\mu)) = \frac{-2}{(n+1)} A_n (\lambda(\mu)).\eqno(184)$$
Together, eqs. (163) and (183) show that
$$A_n (\lambda) = K_n\exp \left[- \int_{\lambda_{0}}^\lambda \left(\frac{4}{\beta(x)} - \frac{2}{x}\right)dx \right]\eqno(185)$$
with $K_{n+1} = -2K_n/(n+1)$, with
$$A_n(\lambda) = \frac{(-2)^n}{n!} A_0 (\lambda)\eqno(186)$$
so that eq. (171) becomes
$$Y(\lambda , L) = \sum_{n=0}^\infty \frac{(-2)^n}{n!} L^n A_0(\lambda),\eqno(187)$$
reproducing eq. (183).

The Lagrange density of massless scalar electrodynamics, given by eq. (128), is characterized by the two couplings $\lambda$ and $e^2$, so the effective potential now takes the form
$$V(\lambda, e^2, \phi, \mu) = \sum_{n=0}^\infty A_n (\lambda, e^2)L^n\phi^4\eqno(188)$$
where again $L = \ln\left(\frac{\lambda\phi^2}{\mu^2}\right)$. Substitution of eq. (188) into the renormalization group equation
$$\left[ \mu \frac{\partial}{\partial\mu} + \beta_\lambda(\lambda,e^2)\frac{\partial}{\partial\lambda} + \beta_{e^2} (\lambda, e^2)\frac{\partial}{\partial e^2} + \gamma(\lambda, e^2)\phi \frac{\partial}{\partial\phi}\right] V = 0\eqno(189)$$
gives rise to an analogue of eq. (171)
$$A_{n+1} = \frac{1}{2(n+1)}\left(\tilde{\beta}_\lambda \frac{\partial}{\partial\lambda}  +\tilde{\beta}_{e^2}\frac{\partial}{\partial e^2}  + 4\tilde{\gamma}\right)A_n\eqno(190)$$
where
$$\tilde{\beta}_\lambda  = \frac{\beta_\lambda}{1 - \gamma - \beta_\lambda/2\lambda},\;\;\;\;
\tilde{\beta}_{e^2}  = \frac{\beta_{e^2}}{1 - \gamma - \beta_\lambda/2\lambda},\;\;\;\;
\tilde{\gamma}  = \frac{\gamma}{1 - \gamma - \beta_\lambda/2\lambda}. \eqno(191)$$

Furthermore, if as in eq. (132),
$$\frac{d}{d\phi}V(\lambda , e^2, \phi = <\phi>, \mu) = 0\eqno(192)$$
then, much as we obtained eq. (177), we find that
$$\sum_{n=0}^\infty \left[ 4A_n (\lambda, e^2) + 2(n + 1) A_{n+1} (\lambda, e^2)  \right]\left(\ln \frac{\lambda<\phi>^2}{\mu^2}  \right)^n <\phi>^3 = 0.\eqno(193)$$
If this is to be  satisfied order by order in $\ln \left(\frac{\lambda<\phi>^2}{\mu^2}\right)$ when $<\phi> \neq 0$, then we see that
$$A_{n+1} (\lambda, e^2) = -\frac{2}{n+1} A_n(\lambda, e^2)  ,\eqno(194)$$
so that a generalization of eq. (183) follows
$$V = \frac{A_0 (\lambda, e^2) \mu^4}{\lambda^2} .\eqno(195)$$

An argument presented in ref. [45] also can be used to establish eq. (195) using only the $n = 0$ contribution to eq. (194) with the renormalization group equation of eq. (190).

So also, the massless $\lambda\phi_3^6$ model can be shown by using the techniques which have led to eq. (183) to have an effective potential which is independent of the background  field $\phi$ and to have a non-analytic dependence on $\lambda$ unless the vacuum expectation value $<\phi>$ vanishes.

Finally we shall apply [47] the renormalization group to examine a massive $\lambda\phi_4^4$ model in which the Lagrange density is
$$L = \frac{1}{2} (\partial_\mu \phi)^2 - \frac{1}{2} m^2\phi^2 - \frac{1}{4!} \lambda \phi^4.\eqno(196)$$
The effective potential $V$ for this model has also been considered by [41-42]. In ref. [41] it was noted that the general form of $V$ is
$$V = \left\lbrace \left(a + \frac{b}{x}\right) + \sum_{\ell = 1}^\infty \lambda^{\ell + 1} \sum_{m=0}^{\ell - 1} x^{m-2} \sum_{n=0}^\ell y^n a_{\ell m n}  \right.\eqno(197)$$
$$\left. + \left(\frac{1-x}{x}\right)^2 \sum_{k=1}^\infty b_k\lambda^k\right\rbrace\phi^4.\nonumber$$
where $a = \frac{-5}{24}$, $b = \frac{1}{4}$ and
$$x = \frac{1}{1 + \frac{2m^2}{\lambda\phi^2}},\;\;\;\;
y = \ln \left(\frac{\lambda\phi^2}{2\mu^2} \frac{1}{x}\right) = \ln\left( 
\frac{m^2 + \frac{\lambda\phi^2}{2}}{\mu^2}\right).\eqno(198)$$
This motivates making an expansion in line with those of eqs. (171) and (188) [47],
$$V(\lambda, x, y, \mu, \phi) = \sum_{n=0}^\infty A_n(\lambda, x)y^n\phi^4.\eqno(199)$$
Substitution of eq. (199) into the renormalization group equation
$$\left(\mu^2\frac{\partial}{\partial\mu^2} + \beta(\lambda)\frac{\partial}{\partial\lambda} + \gamma_m(\lambda) m^2
\frac{\partial}{\partial m^2} + \gamma_\phi (\lambda) \phi^2 \frac{\partial}{\partial\phi^2}  \right)V = 0\eqno(200)$$
results in
$$A_{n+1} (\lambda , x) = \frac{1}{n+1} \left[f(\lambda , x) \frac{\partial}{\partial x} + g(\lambda , x)\frac{\partial}{\partial\lambda} + h(\lambda , x) \right]A_n(\lambda, x)\eqno(201)$$
where
$$f(\lambda , x) = \left(\frac{\beta}{\lambda} - \gamma_m + \gamma_\phi\right)x(1-x)/D(\lambda , x)\nonumber$$
$$g(\lambda, x) = \beta/D(\lambda, x)\nonumber$$
$$h(\lambda , x) = 2\gamma_\phi/D(\lambda, x)\nonumber$$
$(D(\lambda, x) = 1 - \left(\frac{\beta}{\lambda} + \gamma_\phi\right)x - \gamma_m(1-x)$) if the renormalization group equation is satisfied at each order in $y$.

Furthermore, since
$$\frac{dV}{d\phi^2} = \sum_{n=0}^\infty \left[\left(x(1-x) \frac{\partial A_n}{\partial x} + 2A_n\right)y^n + nxA_n y^{n-1}  \right]\phi^2\eqno(202)$$
$V$ has an extremum at $\phi = <\phi>$ if at each order in $y$
$$A_{n+1} (\lambda , x_0) = \frac{1}{n+1} \left[-\frac{2}{x}-(1-x) \frac{\partial}{\partial x} \right]A_n(\lambda, x_0)\eqno(203)$$
provided $<\phi> \neq 0$ ($x_0 = \left(1 + \frac{2m^2}{\lambda<\phi>^2}\right)^{-1}$). Taking eq. (203) to hold for all $x$, then it together with eq. (199) results in
$$V = \sum_{n=0}^\infty \frac{y^n}{n!}\left[- \frac{2}{x} - (1-x)\frac{\partial}{\partial x}  \right]^n A_0(\lambda , x)\phi^4.\eqno(204)$$
Upon setting
$$z = \ln \left(\frac{m^2}{\mu^2}\right) - y = \ln (1-x)\eqno(205)$$
and
$$B_0 (\lambda, z) = \exp \left(-2 \int_{z_0}^z \frac{dt}{1-e^t}\right)A_0(\lambda, x)\eqno(206)$$
eq. (204) becomes
$$V = \exp\left(  2 \int_{z_0}^z\frac{dt}{1-e^t}\right)
\sum_{n=0}^\infty \frac{y^n}{n!} \left(\frac{\partial}{\partial z} \right)^n B_0(\lambda , z)\phi^4 \nonumber$$
$$= \exp \left( -2 \int_z^{z+y} \frac{dt}{1 - e^t}\right) A_0(\lambda, z + y)\phi^4.\eqno(207)$$
By eq. (205), eq. (207) collapses down to an expression for $V$ which is independent of $\phi$, much like eqs. (183) and (195),
$$V = \frac{4(m^2 - \mu^2)^2}{\lambda^2} A_0\left(\lambda, \ln \frac{m^2}{\mu^2}\right).\eqno(208)$$
The function $A_0(\lambda, x)$ satisfies an equation that follows from eqs. (201) and (203) when $n = 0$, 
$$\left(- \frac{2}{x} - (1-x) \frac{\partial}{\partial x}\right)A_0(\lambda, x) = \left(f (\lambda, x) \frac{\partial}{\partial x} + g(\lambda, x)\frac{\partial}{\partial \lambda} + h (\lambda, x)\right)A_0(\lambda, x).\eqno(209)$$

There is an alternate way of deriving eq. (208) that does not require that eq. (202) holds order by order in $y$ when $\phi = <\phi>$. This involves setting $\mu^2$ equal to a value $\mu_0^2$ so that $y = 0$, $\lambda = \lambda_0$ and $x = x_0$; in this case only the $y^0$ contribution to eq. (202) survives,
$$A_1 (\lambda_0, x_0) = \left(- \frac{2}{x} - (1-x) \frac{\partial}{\partial x}\right) A_0 (\lambda_0, x_0).\eqno(210)$$
As the actual value of $\lambda_0$ is contingent upon the unspecified value of the boundary condition on the equation for this running coupling constant, eq. (210) can be taken to be a functional relation between $A_0(\lambda, x)$ and $A_1(\lambda, x)$.

We now follow an approach modeled on the method of characteristics. First we define
$$a_n (\overline{\lambda}(t), \overline{x}(t), t) = \exp\left(\int_{t_0}^t d\tau \;
h(\overline{\lambda}(\tau), \overline{x}(\tau))\right)A_n(\overline{\lambda}(t), \overline{x}(t))\eqno(211)$$
where
$$\frac{d\overline{x}(t)}{dt} = f\left(\overline{\lambda}(t), \overline{x}(t)\right)\;\;\; \left(\overline{x}(t_0) = x\right)\eqno(212)$$
$$\frac{d\overline{\lambda}(t)}{dt} = g\left(\overline{\lambda}(t), \overline{x}(t)\right)\;\;\; \left(\overline{\lambda}(t_0) = \lambda\right).\eqno(213)$$
Eqs. (201), (211-213) together lead to
$$\frac{d}{dt} a_n \left(\overline{\lambda}(t), \overline{x} (t), t\right) = (n+1) a_{n+1} \left(\overline{\lambda}(t), \overline{x} (t), t\right)\eqno(214)$$
Upon taking
$$\overline{y}(t) = \ln\left(\frac{\overline{\lambda}(t)\overline{\phi}^2(t)}{2\overline{\mu}^2(t)}\frac{1}{\overline{x}(t)}  \right)\eqno(215)$$
where
$$\frac{d\ln\overline{\phi}^2(t)}{dt} = \gamma_\phi\left(\overline{\lambda}(t)\right)/D\left(\overline{\lambda} (t), \overline{x}(t)\right)\eqno(216)$$
$$\frac{d\ln\overline{\mu}^2(t)}{dt} = 1/D\left(\overline{\lambda} (t), \overline{x}(t)\right)\eqno(217)$$
$$\left(\overline{\phi}^2(t_0) = \phi^2,\;\;\;\;\overline{\mu}^2(t_0) = \mu^2\right)\nonumber$$
then by eqs. (212, 213, 216, 217) we see that
$$\frac{d\overline{y}(t)}{dt} = -1\;\;\;\; \left(\overline{y}(t_0) = y\right)\eqno(218)$$
so that
$$V = \sum_{n=0}^\infty a_n\left(\overline{\lambda}(t), \overline{x}(t), t\right)\overline{y}^n (t)\phi^4\eqno(219)$$
satisfies
$$\frac{dV(t)}{dt} = 0\;\;\;\; \left(V(t_0) = V(\lambda, x, y, \mu, \phi)\right).\eqno(220)$$
Eqs. (214) and (219) now lead to
$$V(t) = \sum_{n=0}^\infty \frac{\overline{y}^n(t)}{n!} \left(\frac{d}{dt}\right)^n a_0
\left(\overline{\lambda}(t), \overline{x}(t), t\right)\phi^4\nonumber$$
$$= a_0\left(
\left(\overline{\lambda}(t + \overline{y}(t)\right), 
\overline{x}\left(t + \overline{y}(t)\right), t + \overline{y}(t)\right)\phi^4.\eqno(221)$$
We now define
$$\tilde{A}_0 \left(\overline{\lambda}(t), \overline{x}(t),t \right) = \left[\exp \int_{t_0}^t \left(\frac{2}{\overline{x}(\tau)} + h(\overline{\lambda}(\tau), \overline{x}(\tau))\right)d\tau \right]\nonumber$$
$$A_0\left(\overline{\lambda}(t), \overline{x}(t)\right)\eqno(222)$$
which by eqs. (211-213) shows that
$$\tilde{A}_0 \left(
\overline{\lambda}(t_0), \overline{x}(t_0),t_0 \right) = A_0(\lambda, x) = 
a_0\left(\overline{\lambda}(t_0), \overline{x}(t_0),t_0 \right) \eqno(223)$$
while eqs. (209, 212, 213) ensure that
$$\frac{d}{dt} \tilde{A}_0 \left(\overline{\lambda}(t), \overline{x}(t), t \right) = -(1 - \overline{x}(t))\frac{\partial}{\partial\overline{x}(t)} \tilde{A}_0 \left(\overline{\lambda}(t), \overline{x}(t), t\right).\eqno(224)$$
The definitions of eqs. (211) and (222) allow us to rewrite eq. (220) as
$$V(t) =\sum_{n=0}^\infty \frac{\overline{y}^n(t)}{n!} \left(\frac{d}{dt}\right)^n
\left\lbrace\left[ \exp - \int_{t_0}^t \frac{2d\tau}{\overline{x}(\tau)}\right]\tilde{A}_0\left(\overline{\lambda}(t), \overline{x}(t), t\right)\right\rbrace\phi^4\eqno(225)$$
which by eq. (224) becomes
$$= \left[ \exp - \int_{t_0}^t \frac{2d\tau}{\overline{x}(\tau)}\right]\sum_{n=0}^\infty \frac{\overline{y}^n(t)}{n!} 
\left[\frac{-2}{\overline{x}(t)} - (1-\overline{x}(t))\frac{\partial}{\partial\overline{x}(t)}\right]^n
\tilde{A}_0\left(\overline{\lambda}(t), \overline{x}(t), t\right)\phi^4\eqno(226)$$
Just as eq. (204) led to eq. (207), eq. (226) becomes
$$V(t) = \left[\exp - \int_{t_0}^t \frac{2d\tau}{\overline{x}(\tau)}\right]\exp
\left(-2 \int_{\overline{z}(t)}^{\overline{z}(t)+\overline{y}(t)}
\frac{d\tau}{1-e^\tau}\right)
\tilde{A}_0\left(\overline{\lambda}(t), \overline{z}(t) + \overline{y}(t), t\right)\phi^4.\eqno(227)$$
In eq. (227) we can set $t = t_0$; eq. (220) ensures that eq. (208) is recovered.

\noindent
{\bf\large(IX)} It has been demonstrated in this review that the renormalization group is an effective tool for obtaining information about quantum processes beyond what is learned from purely perturbative calculations carried out to finite order. Further applications, such as an examination of the effective potential in massive scalar electrodynamics, are currently being considered.

The author would like to thank his many collaborators, especially M. Ahmady, F. Brandt, F. Chishtie, V. Elias, A. Fariborz, R. Mann, A. Rebhan, T. Sherry and T. Steele.
This work was supported by NSERC. R. Macleod was helpful in the final formulation.

\end{document}